\documentclass[twocolumn,prl,aps]{revtex4}

\usepackage[dvips]{graphicx}

\begin{document}
\title{Gravity compensation in complex plasmas by application of a
temperature gradient}
\author{H.Rothermel, T.Hagl, G.E.Morfill, M.H.Thoma, and H.M.Thomas \\
Centre for Interdisciplinary Plasma Science\\
Max Planck Institut f\"{u}r extraterrestrische Physik, Postfach
1312, D-85741 Garching, Germany\\}

\begin{abstract}
Micron sized particles are suspended or even lifted up in a gas by
thermophoresis. This allows the study of many processes occurring
in strongly coupled complex plasmas at the kinetic level in a
relatively stress-free environment. First results are presented.
The technique is also of interest for technological applications.
\end{abstract}

\maketitle


Dusty or complex plasmas are low temperature, low pressure
plasmas, e.g. glow discharge plasmas, containing microparticles in
addition to ions, electrons and neutral gas. They have a large
variety of important applications in fundamental and applied
physics (see e.g. \cite{bouchoule}). For example, the
microparticles can, under certain conditions, form a crystal
\cite{I, Thomas1, hayashi} which allows the study of the
liquid-solid phase transition at the kinetic level \cite{Thomas}.
Due to the large mass of the microparticles (typically $10^{13}$
times the mass of plasma ions) gravity is an important force. The
normal procedure for complex plasma experiments under gravity
conditions is to suspend the microparticles (which carry a
negative charge, $Q$) in a strong electrostatic field, $E$, i.e.
$mg=QE+\sum_{i} F_{i}$, where the $F_i$ are other forces such as
ion drag. To lowest order, $E$ varies linearly with position in a
sheath of extent and it is easy to calculate that systematic body
force variations from one lattice plane to the next
exceed interparticle forces
for the interesting particle size range above a few $\mu$m. The
systems are thus subject to a considerable amount of stress and
there is abundant stored energy, which affects especially
nonlinear processes such as phase transitions, interface dynamics
and configurational or velocity space instabilities. It is for
this reason that experiments under microgravity were proposed and
are now being conducted \cite{morfill}.

It is the purpose of this paper to demonstrate that for
monodisperse particles thermophoresis can be employed to
counteract gravity, allowing complex plasma experiments to be
performed in the "stress free" environment of the central nearly
field-free plasma of a RF discharge, and to present first
measurements. Results show that large stable systems can be
produced in the charge neutral section of the plasma which exhibit
the "central void" generally observed under 0-g condition. New is
the observed saturation of thermophoretic levitation in the case
of a large particle number combined with low working pressure.
Jellum et al. \cite{jellum} were the first to apply a temperature
gradient and to identify the thermophoretic effect in a dusty
plasma. In order to compensate gravity, however, more
prerequisites have to be fulfilled: a homogeneous temperature
gradient by design, very low energy input to the gas, and the use
of monodisperse particles.

Using elementary kinetic arguments, the thermophoretic force can
be estimated similar to the heat conductivity \cite{reif}. The
latter follows from the energy flux in the presence of a
temperature gradient. Instead of the energy transfer, we consider
the momentum transfer per unit time and area in a plasma. Then the
thermophoretic force $F$ on a microparticle with radius $r_p$ is
given as
\begin {equation}\label{force}
F=-\frac{8}{3}\frac{r_p^{2}}{v}\Lambda \; \frac{dT}{dx},
\end{equation}
where $v=(8kT/\pi m)^{1/2}$ is the average thermal velocity of the
gas atoms with mass $m$ at a gas temperature $T$ and $k$ is the
Boltzmann constant. $\Lambda $ is the coefficient of heat
conductivity and $dT/dx$ the temperature gradient.

As we will see, this formula provides very good results if
measured values or the best theoretical calculations are used for
$\Lambda$. The following result for mono atomic gases has been
given in the literature \cite{gerthsen}
\begin{equation}\label{Lambda}
\Lambda =2.4\> \frac{\eta \> c}{m},
\end{equation}
where $c=3k/2$ is the specific heat per atom and $\eta$ the shear
viscosity of the gas. For the viscosity, we use a result derived
from a precise solution of the transport problem in the case of
hard spheres \cite{reif}
\begin{equation}\label{viscosity}
\eta=0.553\; \frac {\sqrt{mkT}}{\sigma}
\end{equation}
with the gas kinetic cross section $\sigma$ for atomic scattering,
which can be taken from the literature \cite{varney,wutz}. Using
these cross sections, values for the viscosity of noble gases are
obtained (see Table I) in good agreement with measured ones. For
example, the measured viscosity of Argon at $T=300$ K is $\eta =
2.21\cdot 10^{-5}$ Nsm$^{-2}$ compared to $2.16\cdot 10^{-5}$
Nsm$^{-2}$ following from (\ref{viscosity}) together with the
cross section $\sigma = 4.2 \cdot 10^{-19}$ m$^2$ given in Table
I. Combining (\ref{force}), (\ref{Lambda}), and (\ref{viscosity})
results in the thermophoretic force on a microparticle
\begin{equation}\label{force2}
F=-3.33\> \frac{kr_p^{2}}{\sigma}\> \frac{dT}{dx}.
\end{equation}
Like viscosity and heat conduction the thermophoretic force is
pressure independent, as can been seen from (\ref{force2})
containing no pressure dependent parameters. Eq. (\ref{force2}) is
valid only for spherical particles, mono atomic gases and low
pressure where the mean free path is much larger than the particle
radius.

We now compare our result to the literature. First, Waldmann
\cite{waldmann} derived the thermophoretic force using the
Enskog-Chapman method for solving the Boltzmann equation. He
obtained the same expression as (\ref{force}), where, however, the
coefficient is given by 32/15 instead of 8/3, i.e., a
thermophoretic force reduced by about 20\% compared to our result.

In order to compare our result with the one by Havnes et al.
\cite{havnes}, we use the relation $n\lambda=1/(\sqrt{2}\sigma)$
\cite{reif} in (\ref{force2}), where $n$ is the number density of
the gas and $\lambda$ the mean free path of the atoms, leading to
\begin{equation}\label{force3}
F=-4.67\> nk\lambda r_p^{2}\> \frac{dT}{dx}.
\end{equation}
The result obtained by Havnes et al. follows from replacing the
factor 4.67 by 8 in (\ref{force3}), i.e., they found a
thermophoretic force which is larger than ours by almost a factor
of two. Balabanov et al. \cite{balabanov} have used a formula for
the thermophoretic force in a complex DC plasma which agrees with
(\ref{force3}) assuming the ideal gas relation for the pressure
and replacing the coefficient 4.67 by 4.

\begin{table}[htbp]
\caption{\label{sect}Gaskinetic atom-atom cross-sections $\sigma$,
mean free path $\lambda (p)$ at 50~Pa, viscosity and heat
conductivity at 293~K, interaction radii $R=2r=\sqrt{\sigma/\pi}$
from two sources. The thermophoretic force $F$ is compiled for the
actual particle radius $r_p = 1.69$ $\mu$m and temperature
gradient of 1170~K/m. For a given temperature difference, $F$ is
strongest for Helium.}

\begin {center}
\begin{tabular}{l|c|lllllc}
     &Dim.             & He & Ne  & Ar& Kr  &Xe  &Ref.   \\
\hline
$\sigma$ &$10^{-20}$m$^{2}$ &15  & 21   & 42& 49  & 67&\cite{varney} \\
$\lambda (p)$&$10^{-4}$m       &3.52& 2.51 & 1.26& 1.08& 0.79& \\
$\eta$ &$10^{-5}$Nsm$^{-2}$ &1.92&3.06&\bf{2.16}&2.69&2.45& (\ref{viscosity})\\
$\Lambda$ &$10^{-2}$Wm$^{-1}$K$^{-1}$ &14.28&4.56&\bf{1.61}&0.95&0.56&(\ref{Lambda})\\
$R$                  &$10^{-10}$m&2.19&2.59&3.66&3.95&4.62&\cite{varney}\\
$R$                  &$10^{-10}$m&2.18&2.56&3.66&4.14&4.88&\cite{wutz}\\
$F                $ &$10^{-13}$N&10.2&7.31&\bf{3.66}&3.13&2.29&\cite{hayashi}\\
\end{tabular}
\end {center}
\end{table}

\begin{figure}[htbp]
\includegraphics{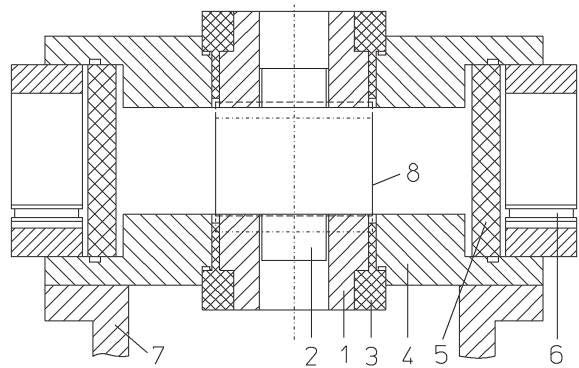}
\caption{\label{fig1}Cross-section through the chamber: (1)
stainless steel electrode, (2) dust dispenser, (3) Macor
insulator, (4) metal plate at chassis potential, (5) window
(welded Vycor body), (6) Peltier element, (7) heat sink, (8-10)
fields of view (thermophoresis on: dashed line, off: dotted
dashed). The distance between the electrodes is 30~mm.}
\end{figure}

{\bf Measurements} were done in a completely symmetrical
RF-excited plasma chamber with a volume of 400 cm$^3$
(Fig.~\ref{fig1}). The gas is heated from below and cooled from
above by Peltier elements. The RF amplitude (with frequency
13.56~MHz) is applied to the electrodes with $180^0$ phase
difference.

\begin{figure}[htbp]
\includegraphics{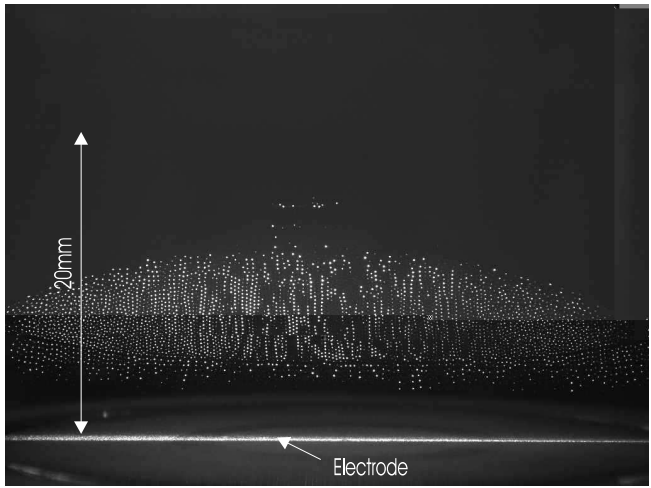}
\caption{\label{fig2} Side view of 3.4$\mu$m microparticles
suspended in a plasma (complex plasma), when no external
temperature gradient is present. The gas is Argon, the pressure
48~Pa, the RF amplitude 45V peak to peak, the discharge power
17~mW. The number of particles is about $10^6$, the field of view
32x43mm. The lowest 5 lattice planes of large interparticle
distance consist of agglomerates.}
\end{figure}

\begin{figure}[htbp]
\includegraphics{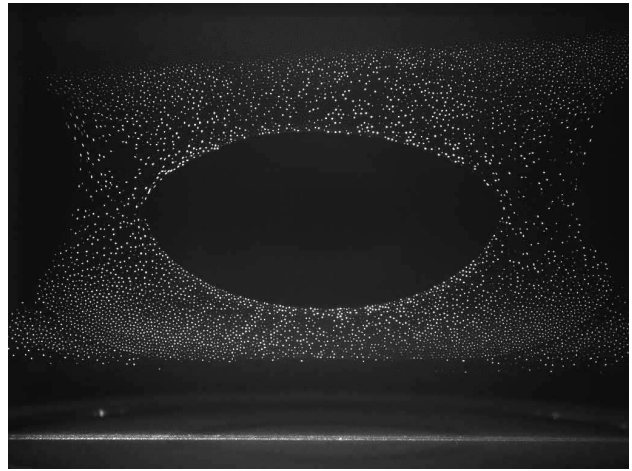}
\caption{\label{fig3}Side view of a complex plasma at quasi 0-g
accomplished by thermophoresis. The temperature gradient applied
from outside is 1170~K/m. The peak to peak amplitude is 82V, the
discharge power 57~mW, the pressure 46~Pa and the number of
particles injected in the order of 1 million.}
\end{figure}

For calibration of the thermophoretic effect the discharge was
operated at only 20~mW. About $10^{4}$ particles were injected,
much less than shown in e.g. Fig.\ref{fig2}. The temperature
difference between the metal plates ((4) in Fig.\ref{fig1}) was
increased until two narrow particle clouds, above and below the
central void (Fig.\ref{fig3}), of equal size were visible. For
Argon gas the equilibrium was reached for temperatures of 54.6~C
at the lower electrode and 19.5~C at the upper electrode. The
temperature difference corresponds to a gradient of
1170~Km$^{-1}$. As monodisperse particles still have a residual
size and weight spectrum, particles observed above the void are
different from those below in that they are about 5\% less heavy.
The 50/50 partition hence calibrates the thermophoretic force by
the average particle weight. For the measured temperature gradient
of 1170~Km$^{-1}$ (\ref{force2}) predicts a thermophoretic force
of $3.66\cdot10^{-13}$ N. Our particles (monodisperse melamine
resin, diameter $3.4 \pm 0.1 \mu$m, density 1510 kg/m$^3$) have a
weight of $3.0\pm 0.3\cdot10^{-13}$ N. Thus, formula
(\ref{force2}) is confirmed within the experimental uncertainties.
We repeated the calibration with Neon and found equilibrium at a
temperature difference of 17.7~K. Thus thermophoresis in Neon is
twice as strong as in Argon as expected from the gas kinetic
cross-section in Table \ref{sect}. We also verified the 50/50
partition as described above for a pressure range of 8 to 62~Pa
and found no pressure dependence in thermophoresis as predicted by
theory.

For comparison and further description of the apparatus we start
with a condition where the Peltier elements and thermophoresis are
not active (Fig. \ref{fig2}). The particle cloud settles down
above the lower electrode due to gravity. One observes that the
particles form a lattice, the so-called plasma crystal. The plasma
glow was filtered away by an interference filter peaked on the
laser wavelength of 686 nm. Measurements taken without filter
indicate the highest particle density congruent with the brightest
glow. There is no convection after the particles have arranged
themselves into the ordered structure. The laser cut is focused to
$\approx$50$\mu$m full width, much narrower than the lattice
spacing. The stringlike structures, caused by the wake potential
of the positive ions streaming to the lower electrode, disappear
in patches, corresponding to local distortions no longer
illuminated. The three dimensional structure of such a system was
analyzed by Zuzic et al. \cite{zuzic}. The RF power measured at
the generator is 170~mW, the peak to peak amplitude measured by a
voltage probe at the electrode is 45V. Both electrodes are shunted
with 3.3 ~k$\Omega$ to chassis. Losses in the unbalanced/balanced
transformer, the matching circuits and in particular the shunt
resistors dictate that only 10\% (17~mW) of the RF power go to the
discharge.

For the thermophoretic measurements the predetermined gradient of
1170 K/m was applied, the plasma was started and the discharge
power was set to a medium value of 20~mW. Particles were injected
until the plasma appeared to be saturated. To accomplish a stable
discharge in the quasi 0-g case, higher RF amplitudes and powers
are required. A survey was measured at three power levels (40, 57,
72 mW) and four pressures (14, 24, 38, 46 Pa).

\begin{figure}[htbp]
\includegraphics{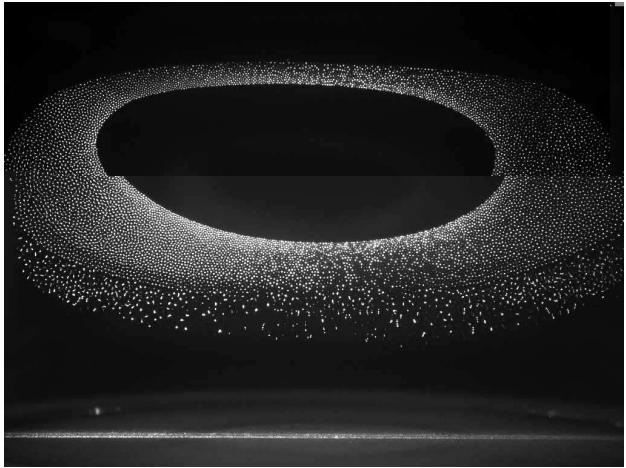}
\caption{\label{fig4}Side view under identical conditions as Fig.
\ref{fig3} with only the gas pressure reduced from 46~Pa to 14~Pa.
As the thermophoretic force becomes saturated particles settle
down below the void. Interesting is a strong enhancement of
particle density behind a very sharp void boundary.}
\end{figure}

Fig. \ref{fig3} and \ref{fig4} are taken out of the survey in
order to visualize the stress-free complex plasma under quasi 0-g
and in order to show a saturation of thermophoresis in the
presence of a large particle number combined with a small working
pressure. Fig. \ref{fig3} shows how the particle cloud fills the
whole discharge volume apart from space charge regions near to the
electrodes and the central region where a void is formed
\cite{morfill}. The reason for no lattice structure in Fig.
\ref{fig3} could be a residual convective motion in the order of
1mm/s which is not visible in a single video frame. The void
boundary appears well defined with a slight enhancement of
particle density behind the boundary. Using $\Lambda$ as given in
table~1 we calculate a total heat flow from the lower to the upper
electrode of 24~mW. On the other hand the RF power dissipated in
the plasma is roughly 57~mW. Most of this power will be dissipated
by ionisation and heating of the electrodes by ion impact. Only a
few mW will heat the neutral gas directly, hence we expect that
the temperature profile applied from outside for thermophoresis
remains reasonably homogeneous for the discharge parameters used,
although a small amount of Joule heating in the center of the chamber 
might contribute to the void formation.
If we compare our experimental condition to the one reported by
Jellum et al. \cite{jellum}, we find from the amplitudes and the
power levels that they drive their discharge at 5 to~10 times the
power density that we use. If we increase the discharge power by
10, we find that the particles are driven out of the bulk plasma
by both the ion drag force and intrinsic thermophoresis.

Obviously thermophoresis will not support an unreasonably large
total mass of particles. A saturation in the thermophoretic force
becomes visible first at low working pressure. Fig. \ref{fig4} is
a measurement under identical conditions as Fig. \ref{fig3},
however, at a working pressure of 14~Pa. The particles settle down
towards the lower electrode. This effect can be reversed by
increasing the pressure again to 46~Pa. Out of a number of
possible mechanisms our data as well as rough estimates identify
radiation cooling (Brattli and Havnes \cite{brattli}) as the best
explanation. By thermal radiation the particles will approach 
everywhere a
temperature roughly equal to the average temperature of the chamber surfaces.
Hence for a large particle density and good thermal
coupling between the particles and the gas the temperature gradient
in the dust region will be reduced and the gradient in the dust free
region will become correspondingly larger.  
A reduced temperature 
gradient in the dust region weakens the thermophoretic force.
Gravity gains the overhand and the dust settles down towards the
lower (heated) electrode as observed. At larger pressure the decreasing
mean free path of the neutral atoms reduces the thermal coupling
between the gas and the particles, causing the reversal of this effect.
A quantitative understanding
of this effect requires detailed investigations beyond the scope of the 
present Letter.

At ambient pressure ($10^{5}$ Pa) gravity driven convection would
obstruct thermophoresis completely. At working pressures between
46 and 14~Pa convection is 3 orders of magnitude weaker. The
driving force of vorticity depends on the temperature difference,
the free fall acceleration and the mass density of the gas. The
driving force hence decreases in proportion to pressure whereas
the counteracting friction on the chamber walls remains unchanged
as the viscosity of gases is pressure independent.

In {\bf conclusion}, 
the thermophoretic levitation of microparticles in a complex plasma,
can give interesting insights into complex plasmas and be used 
as a promising tool complementary to microgravity experiments. 
As an example we discussed 
radiation cooling by the particles, leading to the observed particle 
density dependent reduction of the temperature gradient. Other important
exampels are the investigation of the void and vortices formation, also
observed under microgravity conditions. Furthermore we present
an easy to use relation for the thermophoretic force together with
the relevant gas kinetic data. Technological applications of
thermophoresis could be size selective particle production in a
reactive plasma and surface plating of particles. In semiconductor
production where nanometer sized particles are a byproduct in some
processes, controlled thermophoresis may be used to keep particles
away from the substrate where they can diminish the yield, as already 
suggested by Jellum et al. \cite{jellum}, or to
incorporate them in a controlled way where they are beneficial,
e.g. for amorphous solar cells \cite{rocca}. Controlled
thermophoresis could be interesting for "pick and place" in
context with hybrid integration \cite{pick-place}.


{\bf Acknowledgements}

The work was triggered by a microgravity project funded by: DLR
(BMBF), grant no. 50WB9852. We have to thank Prof. O. Havnes
(University of Troms\o) and our colleagues U. Konopka, K.
Tarantik and M. Zuzic for valuable suggestions and help with the
manuscript.



\begin{thebibliography}{}


\bibitem{bouchoule} A. Bouchoule (Ed.), {\it Dusty Plasmas} (John Wiley,
Chichester 1999).

\bibitem{I} J.H. Chu and Lin I, Phys. Rev. Lett. {\bf72},
4009 (1994).

\bibitem{Thomas1} H.M. Thomas et al., Phys. Rev. Lett. {\bf73},
652 (1994).

\bibitem{hayashi} Y. Hayashi and K. Tachibana,
Jpn. J. Appl. Phys. Part 1 {\bf33}, 804 (1994).

\bibitem{Thomas} H.M. Thomas and G.E. Morfill, Nature {\bf 379}, 806 (1996).

\bibitem{morfill}G.E. Morfill, H.M. Thomas, U. Konopka, H. Rothermel, M. Zuzic, A. Ivlev, and J.
Goree, Phys. Rev. Lett. {\bf83}, 1598 (1999).

\bibitem{jellum} G.M. Jellum, J.E. Daugherty, and D.B. Graves,
J. Appl. Phys. {\bf69}, 6923 (1991).

\bibitem{reif} F. Reif, {\it Fundamentals of Statistical and Thermal Physics}
(McGraw-Hill, New York 1965).

\bibitem{gerthsen} C. Gerthsen, {\it Physik} (Springer, Berlin 1958).

\bibitem{varney} R.N. Varney, Phys. Rev. {\bf88}, 362 (1952).

\bibitem{wutz} M. Wutz, H.Adam, and W. Walcher,
{\it Theorie und Praxis der Vakuumtechnik} (Vieweg, Braunschweig
1981).

\bibitem{waldmann} L. Waldmann, Z. Naturforsch. {\bf14a}, 259 (1959).

\bibitem{havnes} O. Havnes, T. Nitter, V. Tsytovich, G.E. Morfill,
and T. Hartquist, Plasma Sources Sci. Technol. {\bf 3}, 448
(1994).

\bibitem{balabanov} V.V. Balabanov et al., JETP {\bf 92}, 86 (2001).

\bibitem{zuzic}M. Zuzic, A. Ivlev, J. Goree, G.E. Morfill, H.M. Thomas, H. Rothermel, U. Konopka, R.
S\"utterlin, and D.D. Goldbeck, Phys. Rev. Lett. {\bf85}, 4064
(2000).

\bibitem{brattli} A. Brattli and O. Havnes, J. Vac. Sci. Technol A,
{\bf14}, 644 (1996).

\bibitem{rocca}P. Rocca i Cabarroccas, P. Stahel, S. Hamma, and Y. Poissant,
 {\it $2^{nd}$ World conference
and exhibition on photovoltaic solar energy conversion}, (Vienna,
1998), 355.

\bibitem{pick-place}J. Wallace: {\sl Hybrid Integration: Light positions optoelectric
parts}, Laser Focus World, Dec. 2000, 30.

\end{thebibliography}
\end{document}